\documentclass[aps,showpacs]{revtex4}
\usepackage[dvips]{graphicx}
\usepackage{amssymb}
\usepackage{amsmath}
\newcommand{\be}{\begin{equation}}
\newcommand{\ee}{\end{equation}}
\newcommand{\ben}{\begin{eqnarray}}
\newcommand{\een}{\end{eqnarray}}
\newcommand{\bes}{\begin{subequations}}
\newcommand{\ees}{\end{subequations}}
\newcommand{\bb}{\bibitem}

\newcommand{\wt}{\widetilde}

\begin{document}
\title{First-order formalism for scalar field in cosmology}
\author{D. Bazeia, L. Losano, and J.J. Rodrigues}
\affiliation{Departamento de F\'\i sica, Universidade Federal da Para\'\i ba\\
 Caixa Postal 5008, 58051-970 Jo\~ao Pessoa, Para\'\i ba, Brazil}
\date{\today}

\begin{abstract}
We present a general procedure to solve the equations of motion for cosmological models driven by real scalar fields with first-order differential equations. The method seems to have great power, since it works for closed, flat or open space-time, for scalar fields with both standard and tachyonic dynamics. We illustrate the procedure solving several examples which model situations of current interest to modern cosmology.
\end{abstract}

\maketitle

\bigskip
\bigskip
{\bf 1. Introduction}
\\

Dark energy is some necessary but yet very mysterious type of energy. It is introduced to respond for about 3/4 of the physical contents required to fuel the accelerated expansion of the Universe \cite{a1} -- see also \cite{wmap} for more recent investigations. The unusual properties of dark energy defy the interested community, which has introduced a diversity of mechanisms and models to explain its main characteristics -- see, e.g., the recent works \cite{r1}. A lesson to learn from the many investigations is that the possibility to describe properties of the dark energy in cosmological scenarios is related to the presence of scalar fields, which are simple, dynamical and allow for several distinct mechanisms of current interest to modern cosmology. We can also include an exotic fluid which behaves as the Chapligin gas \cite{kmp}. In this case, the fluid has a bizarre equation of state, making the pressure to depend on the negative of the inverse energy density, $p\sim -1/\rho$. The Chapligin fluid has interesting non-standard features \cite{bj}, some of them bridging non-relativistic fluid dynamics behavior into brane-like properties of relativistic Nambu-Goto actions. In cosmology, an interesting property of the Chapligin fluid is that it may interpolate between different phases of the cosmic evolution, from dust-dominated to de Sitter universe, passing through intermediate phase which mix cosmological constant and stiff matter, where $p=\rho,$ and this has inspired many recent investigations.

An important difficulty related to dark energy is intrinsic to cosmology, which is governed by Einstein's equation. Even under the presence of very important assumptions, which lead to the Friedmann-Robertson-Walker (FRW) model, one still gets Einstein's equation in the form of two non-linear differential equations which are in general very hard to solve. In the presence of scalar fields, other equations should be added, the equations of motion for the several degrees of freedom introduced by the scalar fields present in the model. For this reason, investigations concerning the identification of general properties of the equations of motion in the presence of scalar fields have gained increasing importance, as one sees for instance in the recent investigations  \cite{bglm,st}.

In the investigations presented in \cite{bglm}, some of us have shown how to trade the equations of motion to first-order differential equations, for cosmological models of the FRW type, driven by scalar fields with standard or tachyonic dynamics, in closed, flat or open space-time. Some of these results can be seen as generalizations of the Hamilton-Jacobi formulation of inflation which appeared before in Ref.~{\cite{ll}}.
Evidently, the presence of first-order differential equations is important to inject new motivation into the subject. 

The procedure introduced in \cite{bglm} is inspired in the fact that in FRW cosmology driven by a real scalar field $\phi$, both the Hubble parameter and the scalar field depend on time explicitly; however, if we write $t=t(\phi),$ we can turn the Hubble parameter a function of the scalar field,
$H=H(\phi).$ In Ref.~{\cite{bglm}} we have implemented this possibility with the inclusion of another function, $W=W(\phi)$, in the form $H=W(\phi).$ In flat spacetime, this possibility works very nicely, if the scalar field potential can be written as a function of $W$ in a specific manner,
demanded by the equations of motion of the model under investigation. The key point here is that $H=W(\phi)$ can be seen as a first-order differential equation for the scale factor $a=a(t)$ which defines the Hubble parameter $H={\dot a}/a,$ with dot standing for the time derivative.  
The use of $H=W$ and the equation of motion for the acceleration gives another first-order equation, ${\dot\phi}=-W_\phi,$ with $W_\phi$ standing for $dW/d\phi.$ Now, these two first-order equations and the Friedmann equation demand that the potential $V=V(\phi)$ has the form $V=3W^2/2-W_\phi^2/2.$ 
This procedure is similar to the Hamilton-Jacobi formulation presented in Chapter 2 of Ref.~\cite{ll}.

In Ref.~\cite{bglm} we have been able to go beyond flat space-time, writing first-order differential equations for all the three distinct cases of spherical, flat and hyperbolic geometry. These results were later used to extend the first-order formalism to the braneworld scenario of the Randall-Sundrum type \cite{rs}, which is driven by warped geometry with a single extra dimension of infinity extent \cite{abl}. The braneworld side of the problem has led us to establish direct connection with former investigations, both in flat \cite{flat} and curved space-time \cite{f2,celi} -- see also Refs.~\cite{st,bc,cl,bbn}, which introduce several considerations concerning supersymmetry and the first-order formalism for brane and for cosmology. These findings are now used to further explore the first-order formalism put forward in Refs.\cite{bglm,abl}. Below we consider models of FRW cosmology, driven by real scalar field which evolves with standard or tachyonic dynamics. In the case of standard evolution, we focus on the generality of the procedure and we illustrate the method with some examples of current interest to cosmology. For tachyonic dynamics, we also make specific progress, since we extend the result of \cite{bglm}, which was obtained in flat space-time, to the case of spherical or hyperbolic geometry.

\bigskip\
{\bf 2. Standard dynamics}
\\

To make the above reasonings sound, we investigate the general problem which is described by the action
\be\label{model}
S=\int\,d^4x\;{\sqrt{-g}\;\left(-\frac14\,R+{\cal L}(\phi,\partial_\mu\phi)\right)}
\ee
where $R$ and $\phi$ stand for the scalar curvature and scalar field, respectively, and we are using ${4\pi G}=1.$ As usual, we use $g$ to represent the determinant of the metric tensor $g_{\mu\nu},$ which can be obtained by the line element
\be
ds^2=dt^2-a^2(t)\;\left(\frac{dr^2}{1-kr^2}+r^2d\Omega^2\right)
\ee
In this expression, the last term describes the angular portion of the three-dimensional space, $a(t)$ is the scale factor, and $k$ is constant:
$k=1,0,$ or $-1$, for spherical, flat, or hyperbolic geometry, respectively. 

The energy-momentum tensor has the form $T^\mu_{\;\;\nu}=(\rho,-p,-p,-p),$ where $\rho$ and $p$ represent energy density and pressure, respectively, which can be expressed in terms of the scalar field described by the Lagrange density ${\cal L}.$ The above action leads to Einstein's equation, which allows obtaining
\bes\label{fe}
\ben
H^2=\frac23\,\rho - \frac{k}{a^2}
\\
\frac{\ddot a}{a}=-\frac13\,(\rho+3p)
\een
\ees
The equation of motion for the scalar field depends on ${\cal L}(\phi,\partial_\mu\phi);$ for standard dynamics we have 
\ben\label{sm}
{\cal L}=\frac12\partial_\mu\phi\partial^\mu\phi-V(\phi)
\een
and the equation of motion for $\phi$ has the form
\be\label{em1}
\ddot\phi+3H\dot\phi+\frac{dV}{d\phi}=0
\ee 
The energy density and pressure which account for standard dynamics are given by 
\be\label{rp}
\rho=\frac12\dot\phi^2+V,\;\;\;\;\;p=\frac12\dot\phi^2-V
\ee
We use Eqs.~(\ref{rp}) and (\ref{fe}) to write  
\bes\label{em0}
\ben
H^2=\frac13{\dot\phi}^2+\frac23V-\frac{k}{a^2}
\\
{\dot H}=-\dot\phi^2+\frac{k}{a^2}
\een\ees
The set of Eqs.~(\ref{em1}) and (\ref{em0}) constitutes the equations we have to deal with to solve the problem specified by the potential $V(\phi).$

To trade this set of Eqs.~(\ref{em1}) and (\ref{em0}) to first-order differential equations, we follow Refs.~\cite{bglm,abl}. The first step is to make the choice
\be\label{m-g1}
H=W+\alpha k Z
\ee
which leads to the equation
\be\label{m-g2}
\dot\phi=-W_\phi+\beta k Z_\phi
\ee
where $W=W(\phi)$ and $Z=Z(\phi)$ are in principle arbitrary functions of the scalar field, and $\alpha$ and $\beta$ are real parameters. 
These equations are first-order differential equations, and they solve the set of equations (\ref{em1}) and (\ref{em0}) for the potential
\be
V=\frac32\,(W+\alpha k Z)^2+\frac12\,(\beta k Z_\phi-W_\phi)[(3\alpha+2\beta)kZ_\phi+W_\phi]
\ee
and the constraint
\be
W_{\phi\phi}Z_\phi+W_\phi Z_{\phi\phi}-2\alpha k Z Z_\phi-2\beta k Z_\phi Z_{\phi\phi}-2WZ_\phi=0
\ee

We can make the choice $\alpha=0.$ It gives 
\be\label{m-1}
H=W
\ee
and
\be\label{m-2}
\dot\phi=-W_\phi+\beta k Z_\phi
\ee
In this case we get the potential
\be\label{m-3}
V=\frac32\,W^2+\frac12(\beta k Z_\phi-W_\phi)(2\beta k Z_\phi+W_\phi)
\ee
and constraint
\be\label{m-4}
W_{\phi\phi}Z_\phi+W_\phi Z_{\phi\phi}-2\beta k Z_\phi Z_{\phi\phi}-2WZ_\phi=0
\ee
This choice leads us back to the investigations done in Ref.~{\cite{bglm}}, if we change $Z_\phi$ to $Z,$ as we have used in \cite{bglm}. Another possibility is given by $\beta=0,$ which leads to 
\be\label{m-a}
H=W+\alpha k Z
\ee
and
\be\label{m-b}
\dot\phi=-W_\phi
\ee
In this case we get the potential
\be\label{m-c}
V=\frac32\,(W+\alpha k Z)^2-\frac12\,W_\phi(3\alpha k Z_\phi+W_\phi)
\ee
and constraint
\be\label{m-d}
W_{\phi\phi}Z_\phi+W_\phi Z_{\phi\phi}-2\alpha k Z Z_\phi-2WZ_\phi=0
\ee

To investigate the cosmic acceleration, instead of using the deceleration parameter $q=-{\ddot{a}a}/{\dot{a}^{2}}$ we introduce
$\bar q=-q$ as the acceleration parameter, which has the form
\be 
\bar{q}=\frac{\ddot{a}a}{\dot{a}^{2}}=1+\frac{\dot{H}}{H^{2}}
\ee\\ 
We use Eqs.~(\ref{m-g1}) and (\ref{m-g2}) to get
\be
\bar{q}=1+\frac{(W_\phi+\alpha k Z_\phi)(\beta kZ_\phi-W_\phi)}{(W+\alpha k Z)^2}
\ee
For $\alpha=0$ it gives
\be\label{m-5}
\bar{q}=1+\frac{W_\phi(\beta kZ_\phi-W_\phi)}{W^2}
\ee
and for $\beta=0$ it becomes
\be\label{m-e}
\bar{q}=1-\frac{W_\phi(W_\phi+\alpha k Z_\phi)}{(W+\alpha k Z)^2}
\ee
and we see that there are several different possibilities to treat the acceleration.

The choice $\alpha=0$ or $\beta=0$ corresponds to different gauge choice -- see, for instance, Refs.~{\cite{f2,abl,st}}. The case $\beta=0$ will be further investigated in the present work. It is described by the first-order equations (\ref{m-a}) and (\ref{m-b}), and the scalar field model has the potential given by equation (\ref{m-c}), where the functions $W(\phi)$ and $Z(\phi)$ are constrained to obey (\ref{m-d}), and the acceleration parameter
has the form (\ref{m-e}). In this case, we notice that if $Z=Z(\phi)$ is given, we can introduce ${\wt W}=W+\alpha kZ$ in order to rewrite (\ref{m-a}) and (\ref{m-b}) in the form $H={\wt W}$ and $\dot\phi=-{\wt W}_\phi+\alpha k Z_\phi.$ This pair of first-order equations has the very same form of the former pair (\ref{m-1}) and (\ref{m-2}). Thus, choosing the pair (\ref{m-1}) and (\ref{m-2}) or (\ref{m-a}) and (\ref{m-b}) is just a matter of convenience. Evidently, we can also choose the pair (\ref{m-g1}) and (\ref{m-g2}), but in this case one needs to set $\alpha+\beta\neq0.$

We illustrate the subject with some examples. In the case $\beta=0,$ let us first consider the choice $Z=\phi.$ In this case the constraint leads to 
\be
W_{\phi\phi}-2W-2\alpha k\phi=0
\ee
We can solve this constraint with $W=\sinh(A\phi)-\alpha k\phi$, for $A=\pm \sqrt{2}.$ The scalar field has the form
\be
\phi=A\;{\rm arctanh}\left[\frac{B}{A+\alpha k}\tanh\left(\frac{AB}{2}t\right)\right]
\ee
with $B=\sqrt{\alpha^2 k^2 - 2},$ which requires that $\alpha^2 k^2-2>0.$ The potential is given by
\be
V=\frac32 {\rm sech}^{2}(A\phi)-[A\cosh(A\phi)-\alpha k]\left[\frac{A}{2}\cosh(A\phi)+\alpha k\right]
\ee
and now the Hubble parameter gets to the form
\be
H=\frac{2B(A+\alpha k)\tanh\left(\frac{AB}{2}t\right)}{(A+\alpha k)^{2}-B^{2}\tanh^{2}(\frac{AB}{2}t)}
\ee
The cosmic acceleration is given by
\be\label{acce1}
\bar{q}=1+\frac{A}{4}\frac{\bigl[(A+\alpha k)^{2}+2\tanh^{2}\left(\frac{AB}{2}t\right)\bigr]{\rm sech}^{2}\left(\frac{AB}{2}t\right)}{(A+\alpha k)\tanh^{2}\left(\frac{AB}{2}t\right)}
\ee
It is plotted in Fig.~1 for for some specific values of $\alpha$ and $k.$ We see that for $k=-1$ the acceleration changes sign, nicely indicating that
the cosmic evolution may evolves from deceleration to acceleration.  

\vspace{.6cm}
\begin{figure}[!htb]
\includegraphics[scale=0.32, angle=-90]{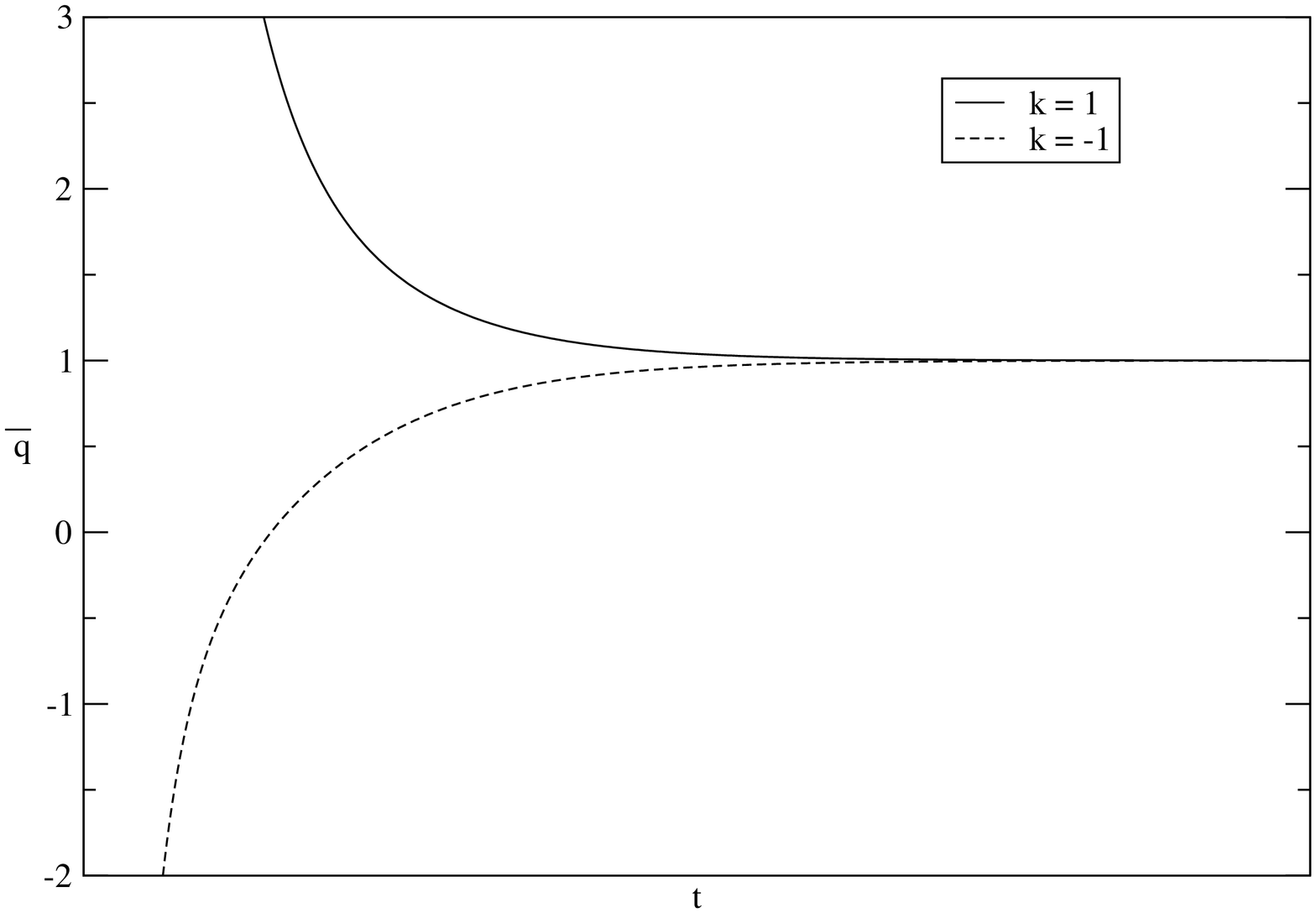}
\caption{Plots of the acceleration (\ref{acce1}) for $\alpha=2,$ and $k=\pm1$.}
\end{figure}

We take the same model, but we now choose $Z=W.$ The constraint equation reduces to 
\be
W_{\phi\phi}-(1+\alpha k)W=0
\ee\\
We can solve this constraint with $W=A\cosh(B\phi)$, com $B=\sqrt{1+\alpha k}.$ The scalar field is given by
\be
\phi=\frac1B \ln\left[\tanh\left(\frac{AB^2}{2}t\right)\right]
\ee
and the potential has the form
\be
V=\frac32 {AB}^{2}\cosh^{2}(B\phi)-\frac12 (1+3\alpha k){AB}^{2}\sinh^{2}(B\phi)
\ee
The Hubble parameter is
\be
H = \frac{AB^{2}}2 \left[\frac{\tanh^{2}\left(\frac{AB^{2}}2 t\right)+1}{\tanh\left(\frac{AB^{2}}2 t\right)}\right]
\ee\\
and the cosmic acceleration gets to the form
\be
\bar{q}= \frac{4\tanh^{2}\left(\frac{AB^2}2 t\right)}{\bigl[\tanh^{2}\left(\frac{AB^2}2 t\right)+1\bigr]^2}
\ee
which is plotted in Fig.~2, showing a new scenario, well distinct from the former one. 
\vspace{.6cm}
\begin{figure}[!htb]
\includegraphics[scale=0.32, angle=-90]{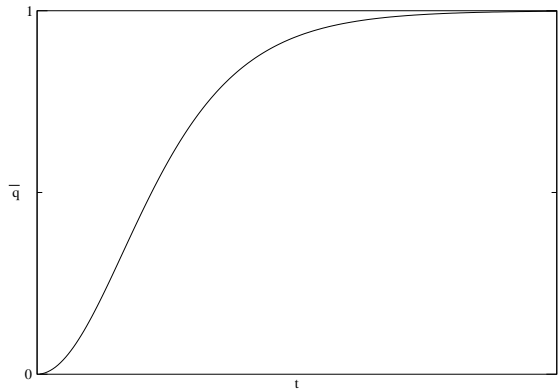}
\caption{Plots of the cosmic acceleration for $A=\alpha k=1.$}
\end{figure}

\bigskip\
{\bf 3. Tachyonic dynamics}
\\

We now turn attention to the case of tachyonic dynamics. The importance of tachyon in cosmology is inspired in String Theory, and can be seen by the recent works \cite{t1}. In this case, the Lagrange density is given by
\be\label{td}
{\cal L}_{t}=-V(\phi)\sqrt{1-\partial_{\mu}\phi\partial^{\mu}\phi}
\ee
The scalar field now evolves differently, according to
\be
\ddot{\phi}+(1-{\dot{\phi}}^{2})\left(3H\dot{\phi}+\frac{V_{\phi}}{V}\right)=0
\ee \\
and the energy density $\rho_{t}$ and pressure $p_{t}$ are changed to
\be
\rho_{t}=V\frac1{\sqrt{1-{\dot{\phi}}^{2}}} \;\;\;\;\;\;\; p_{t}=-V\sqrt{1-{\dot{\phi}}^{2}}
\ee
We use Friedmann's equation and the equation for the acceleration (\ref{fe}) to write, in the presence of curvature
\bes
\ben
H^2=\frac23 \frac{V}{\sqrt{1-{\dot{\phi}}^{2}}}- \frac{k}{a^2}
\\
\dot{H}=-\frac{{\dot{\phi}}^{2}}{\sqrt{1-{\dot{\phi}}^{2}}}V + \frac{k}{a^2}
\een
\ees

We extend the first-order formalism of Ref.~\cite{bglm} to tachyonic dynamics in the presence of curvature as follows. We consider
\be\label{tacnew}
H=\sqrt{W^2-\alpha kZ^2}
\ee
This choice allows writing the equation 
\be
\dot{\phi}=-\frac23 \frac{W_{\phi}}{W\sqrt{(W^{2}-\alpha kZ^2)}}
\ee
In this case, the potential has the form
\be\label{tp}
V=\frac32\,W^2\left[1-\frac49\frac{W^2_\phi/W^2}{W^2-\alpha k Z^2}\right]^{1/2}
\ee
and the constraint becomes
\be
3W^{3}Z-3\alpha kWZ^{3}-2W_{\phi}Z_{\phi}=0
\ee\label{cons}
The acceleration parameter is given by
\be
\bar{q}_t=1-\frac23\,\frac{W_{\phi}}{W}\,\frac{WW_{\phi}-\alpha k ZZ_{\phi}}{(W^{2}-\alpha kZ^2)^2}
\ee
In the slow-roll approximation \cite{ll92}, one uses that $\ddot\phi$ and $\dot\phi^2$ are very small. Here it makes no sense to neglect $\ddot\phi,$
but if we consider $\dot\phi^2$ very small, we can rewrite the potential (\ref{tp}) in the much simpler form
\be
V=\frac32W^{2}-\frac13\frac{W^2_\phi}{W^{2}-\alpha kZ^2}
\ee
and this is the result we could obtain if one uses (\ref{tacnew}), for scalar field evolving in accord to the standard dynamics. This result is very natural: in the tachyonic Lagrange density (\ref{td}), if we expand the square root up to first order in ${\partial_\mu\phi\partial^\mu\phi},$ we get to standard dynamics.

To illustrate the general situation, let us consider some examples. We firstly notice that the case $Z=1$ is trivial, and so we consider $Z=W.$ This case leads us to the constraint 
\be
3(1-\alpha k)W^{4}-2{W_{\phi}}^{2}=0
\ee
which is solved by
\be
W=\pm\frac{\sqrt2}{\sqrt{3(1-\alpha k)}\,\phi}
\ee
In this case we get 
\be
\phi=\pm\sqrt{\frac23}t
\ee
and the potential 
\be
V=\frac{\sqrt3}{3}\left(\frac{1}{1-\alpha k}\right)\frac1{\phi^2}
\ee
The Hubble parameter is $H=1/t,$ and this leads to $\bar{q}=0$.

Another possibility is $Z=W^2.$ Here we get the constraint
\be 
3(1-\alpha kW^{2})W^{4}-4{W_{\phi}}^{2}=0
\ee
which is solved by 
\be 
W=\pm\frac2{\sqrt{4\alpha k +3\phi^2}}
\ee
This gives 
\be 
\phi=\frac{\sqrt3}{3}\,t
\ee
and
\be
V=\frac{2\sqrt6}{4\alpha k + 3\phi^2}
\ee
The Hubble parameter is 
\be
H=\frac{2t}{4\alpha k+t^2}
\ee
and now we get 
\be\label{acce}
\bar{q}=\frac{4\alpha k+t^2}{2t^2}
\ee
We plot this acceleration parameter $\bar q$ in Fig.~3. The new scenarios show that for tachyons the acceleration goes asymptotically to $1/2,$
but the cosmic evolution is very similar to the case shown in Fig.~1.

\vspace{.6cm}
\begin{figure}[!htb]
\includegraphics[scale=0.32, angle=-90]{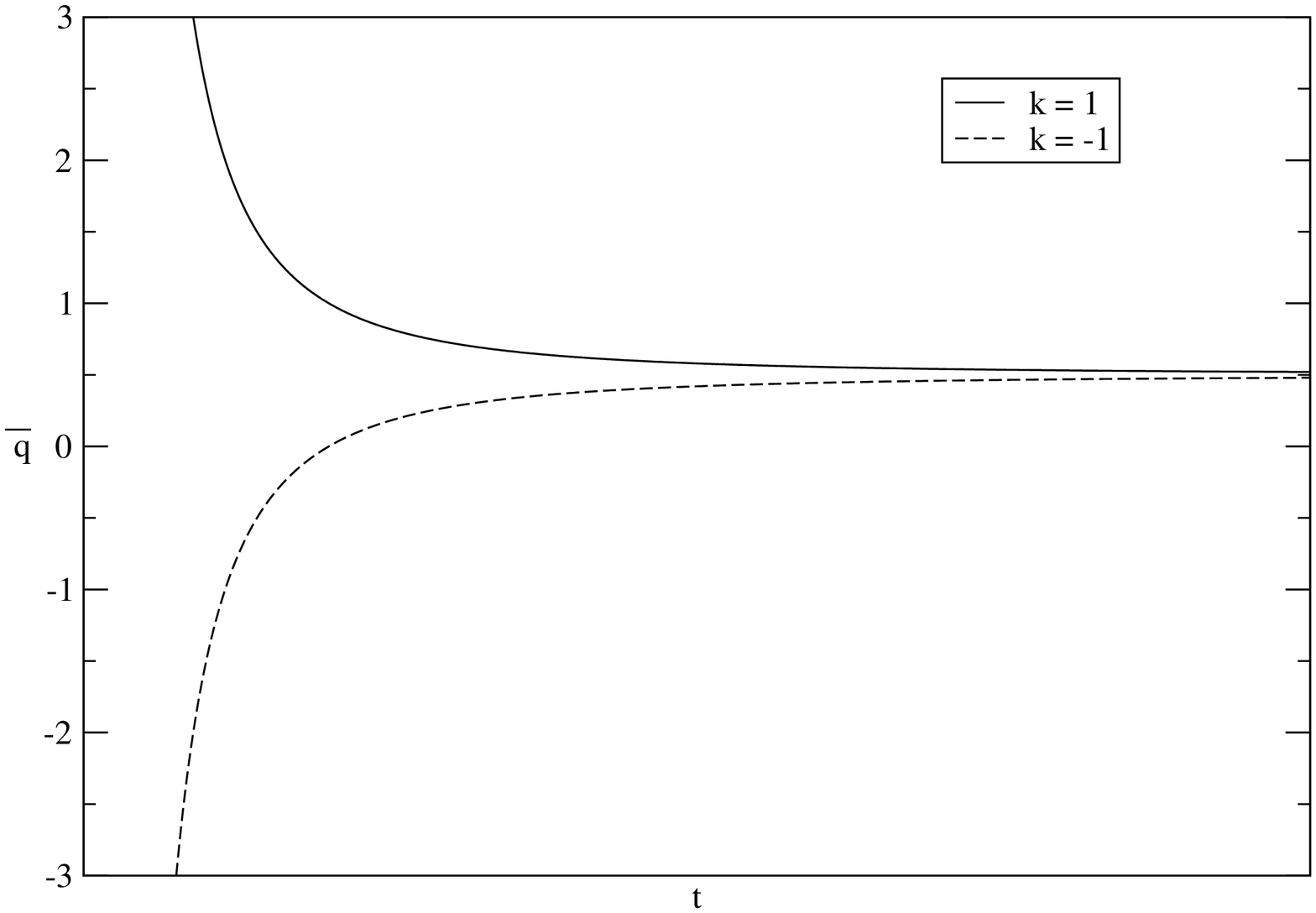}
\caption{Plots of the cosmic acceleration (\ref{acce}) for $\alpha=1,$ and $k=\pm1.$}
\end{figure}

\bigskip\
{\bf 4. Summary}
\\

We summarize this work recalling that we have extended the first-order formalism introduced in Ref.~\cite{bglm} to describe FRW models, driven by a single real scalar field with either standard or tachyonic dynamics for generic spherical, flat or hyperbolic spatial geometry. The importance of the procedure is related not only to the improvement of the precess of finding explicit solution, but also to the opening of another route, in which we can very fast and directly write the Hubble parameter $H$ once $Z=Z(\phi)$ and $W=W(\phi)$ are given. As we have shown, the present investigations are of direct interest to modern cosmology, since they seem to open several distinct possibilities of investigation. In particular, in the case of tachyonic dynamics, we have built the full formalism, including the cases of closed and open geometries, which were not given in the former work \cite{bglm}.
The main results show that we can use standard or tachyonic dynamics to find interesting cosmic evolution, including the case where the cosmic acceleration changes sign, evolving from deceleration to acceleration. 

The interest in the subject will certainly broaden with the extension of the method to the case of several fields, since in this case we could find models in which one field can be used to affect the behavior of other fields, unveiling the possibility to control a given phase and to link different phases of the cosmic evolution. Another possibility is related to the inclusion of the Chapligin \cite{kmp} and other fluid contents, such as dust, radiation and stiff matter. These issues are now under investigation, and we hope to report on them in the near future.
\bigskip

{\bf Acknowledgements}
\\

The authors would like to thank Francisco Brito and Roberto Menezes for discussions, and CAPES, CNPq, and PRONEX/CNPq/FAPESQ for partial support.


\end{document}